\documentclass[twocolumn,prb,amsmath,showpacs,citeautoscript]{revtex4}
\def\etal{{ \it et al. }}
\usepackage{epsfig}
\newcommand{\rb}[1]{\raisebox{1.5ex}[-1.5ex]{#1}}   
\begin{document}
\title{Pressure-dependence of electron-phonon coupling and the 
superconducting phase in hcp Fe - a linear response study} 
\author{S. K. Bose}
\email[on leave from Brock University, St. Catharines, Ontario, 
CANADA L2S 3A1: ]{bose@newton.physics.brocku.ca}
\author{O. V. Dolgov}
\author{J. Kortus}
\author{O. Jepsen}
\author{O. K. Andersen} 
\affiliation{Max-Planck-Institute for Solid State Research, 
Heisenbergstr.\ 1, 70569 Stuttgart, Germany} 

\begin{abstract} 
A recent experiment by Shimizu \etal (Ref. \onlinecite{nature1})
has provided  evidence of a superconducting phase in hcp Fe under pressure. 
To study the pressure-dependence of this superconducting phase  
we have  calculated the phonon frequencies and the electron-phonon coupling 
in hcp Fe as a function of the lattice parameter, using the
linear response (LR) scheme and the full potential linear muffin-tin orbital (FP-LMTO)
method.  Calculated phonon spectra and the Eliashberg functions
$\alpha^2 F$ indicate that conventional $s$-wave electron-phonon coupling 
can  definitely account for the appearance of the superconducting phase in hcp Fe. 
However, the observed change in the transition temperature with increasing 
pressure is far too rapid compared with the calculated results.
For comparison with the linear response results, we have  computed the electron-phonon
coupling  also by using the rigid muffin-tin (RMT) approximation. 
From both the LR and the RMT results 
it appears that electron-phonon interaction alone cannot explain 
the small range of volume  over which  superconductivity is observed.  
It is shown that ferromagnetic/antiferromagnetic spin fluctuations as well as
scattering from magnetic impurities (spin-ordered clusters) can 
account for the  observed values of the transition temperatures but cannot substantially
improve the agreeemnt between the calculated and observed presure/volume range of
the superconducting phase. 
A simplified treatment of $p$-wave pairing leads to extremely small ($\leq 10^{-2}$ K)
transition temperatures. Thus our calculations seem to rule out both $s$- and $p$- wave
superconductivity in hcp Fe.
\end{abstract} 

\maketitle 
\section{Introduction}

Recently Shimizu \etal \cite{nature1} (see also Ref.\onlinecite{jaccard})
 have reported  resistivity and magnetization measurements
on Fe samples under pressure, and identified a superconducting phase 
characterized by   the Meissner effect and the vanishing of the  resistivity  above a pressure of 15 GPa.
 At this pressure the  stable crystal sructure of Fe is known to be hcp.
Both the hcp phase and superconductivity in Fe under pressure are results that can be
expected on theoretical grounds.  Stability
of bcc, fcc and hcp crystal structures as a function of canonical $d$-band filling was discussed
some time back
by Pettifor \cite{pettifor} and Andersen \etal \cite{physica77,varenna}.  These authors showed that
without ferromagnetism the ground state of Fe would be hcp, just as for its nonmagnetic and isoelectronic
4$d$ and 5$d$ counterparts, Ru and Os. For Fe the bcc structure is stabilized only via ferromagnetism.
 In the ferromagnetic bcc  state both the atomic volume and compressibility of Fe are anomalously 
 large\cite{gschneider}. Application of a moderate pressure  results in  
 bcc  to hcp  martensitic transformation  and loss of  
ferromagnetism\cite{physica77}. Both Ru and Os are superconducting at low
temperatures. Thus superconductivity in hcp Fe is  hardly surprising. 

What differentiates hcp Fe from
Ru and Os is the presence of spin fluctuations. Both ferromagnetic and antiferromagnetic spin
fluctuations are known to suppress superconductivity mediated via $s$-wave electron-phonon
coupling. A notable example, where ferromagnetic spin fluctuations (paramagnons) are believed
to suppress superconductivity completely, is fcc Pd. A large density of states (DOS) at the
Fermi level, $N(0)$, in fcc Pd causes a large Stoner-enhanced paramagnetic susceptibility, leading to
strong ferromagnetic spin fluctuations. 
 Disorder-induced superconductivity
 in fcc Pd, due mainly to the reduction in $N(0)$ and therefore  in  spin fluctuations, has been
claimed experimentally  as well as discussed theoretically \cite{stritzker,appel,bose}.
A similar effect could conceivably be achieved in fcc Pd under pressure, but is yet to be observed.
The case for hcp Fe is somewhat different, since it is believed to be close to antiferromagnetic\cite{SN}
   or complex magnetic\cite{cohen} instability. It was noted by Wohlfarth\cite{wohlfarth} that
at the lowest pressures ($\sim$10 GPa) at which hcp Fe is stable, it should be close to
an antiferromagnetic instability. He also suggested that the antiferromagnetic spin fluctuations
might not be strong enough to suppress superconductivity in hcp Fe, particularly at
elevated pressures, where reduction in $N(0)$ would cause spin fluctuations to eventually disappear.
Antiferromagnetic spin fluctuations suppress $s$-wave superconductivity, while contributing to
$p/d$-wave superconductivity. At present experimental evidence regarding the type of superconductivity
($s$-wave or otherwise) in hcp Fe is lacking.

One can estimate the  $T_c$ in hcp Fe by using  simple scaling arguments
and the observed superconducting transition temperature $T_c$ of Ru (0.5 K) or Os (0.7 K)
at normal pressure. 
Let us ignore spin fluctuations and consider the McMillan expression: 
\begin{equation}
T_{c}=\frac{\Theta_D}{1.45}\exp \left\{ -\frac{1.04 \left(1+\lambda _{ph}\right)}{%
\lambda _{ph}-\mu ^{\ast }(1+0.62\lambda _{ph})}\right\} ,  \label{mcm}
\end{equation}
where $\Theta_D$ is the Debye temperature, $\mu^{\ast}$ is the
Coulomb pseudopotential, and $\lambda_{ph}$ is the electron-phonon coupling constant, given by
$\lambda_{ph} = N(0)\langle I^2 \rangle/M\langle\omega^2\rangle$. 
Considering $\mu^{\ast}$ = 0.1,   
 we get $\lambda_{ph} = 0.32$ for
Ru ($\Theta_D$ = 600 K, see Ref.{ \onlinecite{gschneider}}).  
 To estimate
 $\lambda_{ph}$ for hcp Fe, we  assume that the
 mean square electron-phonon(ion) matrix element $\langle I^2 \rangle$ and 
  the effective spring constant $M\langle\omega^2\rangle$  are nearly the same 
 as in hcp Ru. 
  The average phonon frequency, and thus $\Theta_D$,   should then
 scale as the inverse square root  of the ratio of the atomic masses, and $\lambda_{ph}$
should scale according to the ratio of $N(0)$.
The quantity $N(0)$ can be easily calculated for elemental solids. However, we only need to
estimate the  ratio of this quantity  between Ru and Fe. We  can start by
 assuming that $N(0)$ is proportional to the inverse
$d$-band width, which can be estimated from the potential parameters of the
 LMTO-ASA (atomic sphere approximation)  method \cite{varenna}.  
 Here the band width is proportinal to the potential parameter
$\Delta$. Both $\Delta$ and its volume derivative have
already been tabulated\cite{varenna1} for a large number of elemental solids. 
From the  $d$-orbital values of the parameter $\Delta$,
$N(0)_{Fe}/N(0)_{Ru} \simeq \Delta_{Ru}/\Delta_{Fe}$ = 539/280 = 1.925. 
 This would give $\lambda_{ph}$ =0.62  for Fe, resulting in
 a   transition temperature of  $\sim$17 K for hcp Fe. 
If we use published values of $N(0)$,
 for hcp Fe\cite{mazin1} (corresponding to a pressure $P \sim$ 10 GPa) and 
for Ru\cite{ove} (corresponding to normal prssure), then $N(0)_{Fe}/N(0)_{Ru}$ =
20.8/11.8 = 1.76, and $\lambda_{ph}$=0.56
for hcp Fe. 
 This ratio
 yields an  improved value $T_c(Fe)$ = 12 K.
Using the measured value of $\Theta_D$ in hcp Fe ($\sim$ 500 K at $\sim$ 10 GPa, 
see Ref. {\onlinecite{recent-dos}})  gives a value\footnote{The values quoted for $N(0)_{Fe}$\cite{mazin1} and $N(0)_{Ru}$\cite{ove}
are obtained via different methods: LAPW\cite{mazin1} and LMTO\cite{ove}. Using the
same method to calculate both DOSs would result in a ratio that is 10\% smaller,
giving $\lambda_{ph}$ =0.5 for Fe, and a $T_c \sim$ 5 K.} of  7.6 K.

For a quick estimate of the pressure-dependence of $T_c$ we use a simplified version of Eq.\ref{mcm}:
\begin{equation}
\label{simpleTc}
T_c = \frac{\Theta_D}{1.45} \exp \left(\frac{1}{\lambda^{\prime}}\right)\;, \lambda^{\prime}
= \lambda_{ph} - \mu^{\ast}\;,
\end{equation}
 and resort to the tabulated values of the
 logarithmic derivative of the potential parameter $\Delta$ with respect to atomic sphere radius, $s$ 
 (Ref. {\onlinecite{varenna1}}). Neglecting
the volume(pressure) dependence of the quantities $\langle I^2 \rangle$ and $\mu^{\ast}$ in 
Eq.\ref{simpleTc}, we obtain, for the logarithmic derivative of $T_c$ with respect to the system
volume $V$:
\begin{equation}
\frac{d\ln T_c}{d\ln V} = -\gamma_G \left(1-\frac{2}{\lambda_{ph}}\right)  
 + \frac{1}{\lambda_{ph}} \frac{d\ln N(0)}{d\ln V}\;,
\end{equation}
where  $\gamma_G$ is the Gr\"{u}neisen parameter. We have used the approximations:
\begin{equation}
 \gamma_G =- \frac{d\ln\Theta_D}{d\ln V} \approx
-1/2 \frac{d\ln \langle\omega^2\rangle}{d\ln V}\;.
\end{equation}
With the assumption $N(0) \sim \Delta^{-1}$, where $\Delta$ is the $d$-orbital band width parameter
in LMTO-ASA, $d\ln N(0)/d\ln V = -(1/3)d\ln\Delta/d\ln s$. For the $d$-orbitals of 
 Fe $d\ln\Delta/d\ln s$ = -4.6.
  From the reported value\cite{recent-dos} of $\gamma_G = 1.5$ in hcp Fe,
we obtain a value $\frac{d\ln T_c}{d\ln V} =6.6$ for hcp Fe. The zero pressure bulk modulus
in hcp Fe is 165 GPa \cite{mao}. The initial (low pressure) logarithmic derivative of $T_c$ in hcp
Fe should thus be close to -6.6/165 (GPa)$^{-1}$ = - 4\% /GPa.
 
Exercises such as the one outlined above are useful in obtaining order of
magnitude estimates and in understanding the trend from one element to the
next. However, 
 quantitative  agreement with experimental results might  be missing.
 According to the study by Shimizu \etal \cite{nature1} superconductivity in hcp Fe 
appears at around 15 GPa, slightly above the pressure at which the bcc-hcp 
transition takes place. The transition temperature grows slowly from below 1 K to 
about 2 K at $\sim$22 GPa and then decreases steadily, with superconductivity vanishing 
beyond 30 GPa\cite{footnote1}. The rate of decrease of $T_c$ is too rapid compared with the
estimate derived above.
In order to reproduce the initial increase of $T_c$ with pressure, as observed in the experiment,
it would be necessary to include the spin fluctuation effects  and  possible 
  volume dependence of the matrix element $\langle I^2 \rangle$. With 
inceasing pressure, spin fluctuations are expected to diminish, causing $T_c$ to rise.
The electron phonon matrix element may also increase with pressure, as the nearest neighbor
distances become shorter. It would  thus be 
 of interest to examine to what 
extent the observed results can be explained via  a rigorous {\it ab initio} calculation. 
To this end we  have used the full-potential linear muffin-tin orbitals linear 
response (FP-LMTO-LR) scheme developed by Savrasov \cite{savrasov1,savrasov2} 
to calculate the phonon frequencies and the electron-phonon coupling 
in hcp Fe as a function of pressure. The Eliashberg equations  
\cite{allen-mitro}, in their isotropic  Fermi surface averaged form,  are used to study the 
pressure-dependence of the transition temperature $T_c$, and the superconducting gap $\Delta$. 
 Effects of both  ferromagnetic and 
antiferromagnetic spin fluctuations and  the effects of scattering from magnetic impurities 
 are  explored to accommodate the experimental data as best as possible.
 We also
present a  simplified treatment of
$p$-wave pairing in hcp Fe. So far two other theoretical calculations,
related to superconductivity in hcp Fe and its pressure dependence, have appeared\cite{mazin1,jarlborg}. 
  Our work differs from these   publications \cite{mazin1,jarlborg}
 in asmuch as it  presents  a more rigorous first-principles calculation of the phonons
and the electron-phonon coupling as a function of the  lattice parameter in hcp Fe.

\section{Electronic Structure}
 There is
considerable experimental evidence that at room temperature
the martensitic transition from the bcc to the  hcp phase in  iron 
 takes place  at a pressure of 10-15 GPa \cite{jephoat,bassett,taylor}.
Recently Ekman \etal\cite{ekman} have studied this phase transition using  the 
full-potential linear augmented plane wave
(FP-LAPW) total energy method. Their study indicates a first order ferromagnetic bcc to nonmagnetic hcp
transition. 
These authors carried out spin-polarized density functional
calculations using the generalized-gradient approximation (GGA) of Perdew  and Wang (GGA1) \cite{PW1}.
Steinle-Neumann \etal \cite{SN}, using a later version of the GGA by Perdew, Burke, and
Ernzerhofer (GGA2) \cite{PW2}, find  
an antiferromagnetic ground state for hcp Fe and show that this version of the GGA better reproduces
the observed elastic properties of hcp Fe under pressure. The possibility of noncollinear
magnetism in hcp Fe below 50 GPa has also been suggested \cite{cohen}. These  results are at variance with
the M\"{o}ssbauer study of hcp Fe under pressure, which has failed to reveal any local magnetic 
moment \cite{cort,taylor}. The possibility remains that hcp Fe stays close to antiferromagnetic
or complex magnetic
instability. 
 In this work we  assume a nonmagnetic phase for hcp Fe under pressure, and present
results  that were obtained by using GGA1 \cite{PW1}.  
Our calculations, using various forms of GGA, show that the nature of the ground state at various
lattice parameters depends very much on the exchange correlation potential and the
$c/a$ ratio. In addition, the energy differences between  nonmagnetic, ferromagnetic,
and  antiferromagnetic
(AFM I and AFM II configurations\cite{SN} ) states are often small (almost within the
 errors of the method), as well as dependent on technical details,
 such as the number of {\bf k}-points in the
irreducible Brillouin zone and method of BZ integration (see further discussion in section V).

\begin{figure}
\resizebox{!}{2.5in}{\includegraphics{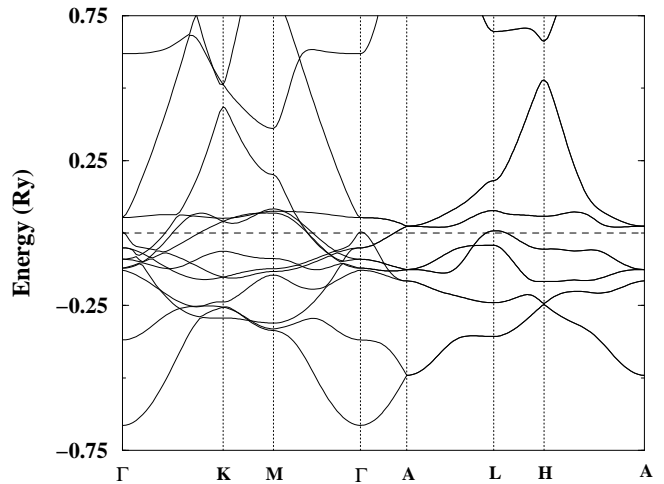}}
\caption[]{ FP-LMTO energy bands in hcp Fe for the ideal c/a value ($\sqrt{8/3}$),
 and a= 4.6 a.u. The horizontal line
shows the position of the Fermi level, chosen as the zero of energy.} 
\label{fig1}
\end{figure}

According to
Ekman \etal \cite{ekman} the bcc to hcp transition leads to  a phase with a $c/a$ ratio of 1.57. 
Our LMTO-ASA  calculations  yield a  smaller (by $\sim$ 0.8 mRy)  hcp ground state energy for $c/a=1.57$
than for the ideal close packing value.  In the FP-LMTO calculations the difference in the ground state energies for the two $c/a$ values is smaller than 0.2 mRy. 
 Previous theoretical studies for iron indicate a very small dependence of the total energy
on the $c/a$ ratio\cite{soderland,stixrude} and in addition, 
the $c/a$ ratio is likely to change with pressure.  
We have  thus adopted the simplest option, as in Ref. \onlinecite{mazin1},
 and carried out all our calculations with
the ideal close-packing case:  $c/a = \sqrt{8/3}$. The electronic structure was
calculated using Savrasov's FP-LMTO code \cite{savrasov-el} with a triple-$\kappa$ $spd$ LMTO basis for the
valence bands. $3s-$ and $3p$-semicore states were treated as valence states in separate energy windows.
The charge densities and potentials were represented by spherical harmonics with $l\leq 6$ inside the 
nonoverlapping MT spheres and by plane waves with energies $\leq$ 141 Ry in the interstitial region.
Brillouin zone (BZ) integrations were performed with the full-cell tetrahedron method \cite{peter} using 793
{\bf k}-points in the irreducible zone.
 Band structures of hcp Fe obtained via the FP-LMTO method for all lattice parameters considered are in
good agreement with the LMTO-ASA bands. FP-LMTO bands
 for the ideal c/a ratio and lattice parameter of 4.6 a.u. are  shown in Fig. \ref{fig1}.

Table \ref{table1}
 shows the lattice parameters used in our calculations together with the atomic volumes and  some 
calculated  properties. 
 Pressure ($-\partial E/\partial V$) and bulk modulus were  calculated by fitting the energy vs. lattice parameter curve to the 
generalized Birch-Murnaghan equation of state \cite{birch,murnaghan}. 
The equilibrium atomic volume and bulk modulus, 69.4 a.u. and 290 GPa, compare well with the values
obtained by the FP-LAPW calculations of Ekman \etal \cite{ekman} (68.94 a.u. and 263 GPa). 
Calculated pressure and bulk modulus values become progressively less reliable away from the equilibrium volume. 
In order to calculate the Stoner parameter $I$ we introduced a small splitting in the self-consistent 
paramagnetic bands  by adding  small up and downward shifts to the band-center parameter $C$ in the
LMTO-ASA method. After making the atom self-consistent the Stoner parameter $I$ was calculated 
from the induced magnetic moment per atom $\mu$, assuming proportionality between band-splitting and the
Stoner parameter: 
\begin{equation}
I = \sum_{l}I_l\delta_l;\; I_l = (C_l^{\uparrow} - C_l^{\downarrow})/\mu ;\; \delta_l = N_l(0)/N(0)\;,
\end{equation} 
where the arrows indicate spin-up and down states and $N_l(0)$ and $N(0)$ are the $l$-partial
and total DOSs at the Fermi level, respectively. This method yields almost the same 
(pressure-independent) value as that obtained by Mazin {\it et al.} \cite{mazin1}. 
 Both our method and that used in Ref. \onlinecite{mazin1} can be called fixed spin-moment method,
except that Mazin {\it et al.} derived $I$ from the second derivative of 
the total energy with respect to the spin-moment.
\begin{table}
\caption{FP-LMTO results for nonmagnetic hcp Fe for the ideal $c/a$ ratio: a= lattice parameter (a.u.),
$V_0$ = volume per atom (a.u.), $P$=pressure (GPa), $B$=Bulk Modulus (GPa), $N(0)$= DOS at the Fermi level
(states/(Ry  atom spin)), $I$= Stoner parameter (Ry/atom).}
\label{table1}
\begin{ruledtabular}
\begin{tabular}{lcccccc}
a  & 4.0 & 4.2 & 4.4 & 4.5 & 4.6 & 4.7\\
$V_0$  & 45.25 & 52.39 & 60.23 & 64.44 & 68.83 & 73.41\\
$P$  & 350 & 162 & 56 & 26 & 2.3 & -14 \\
$B$  & 1695 & 970 & 550 & 410 & 300 &  221\\
$N(0)$  & 4.77 & 5.79  & 7.05  & 7.80  & 8.59  & 9.46 \\
$I$  & 0.074 & 0.074 & 0.073 & 0.073 & 0.073 & 0.075 \\
$1/(1-IN)$ &  1.54 & 1.75 & 2.06 & 2.32 & 2.68 & 3.44\\
\end{tabular}
\end{ruledtabular}
\end{table}

\section{Lattice vibrations and electron-phonon coupling}

\begin{figure}
\resizebox{!}{2.5in}{\includegraphics{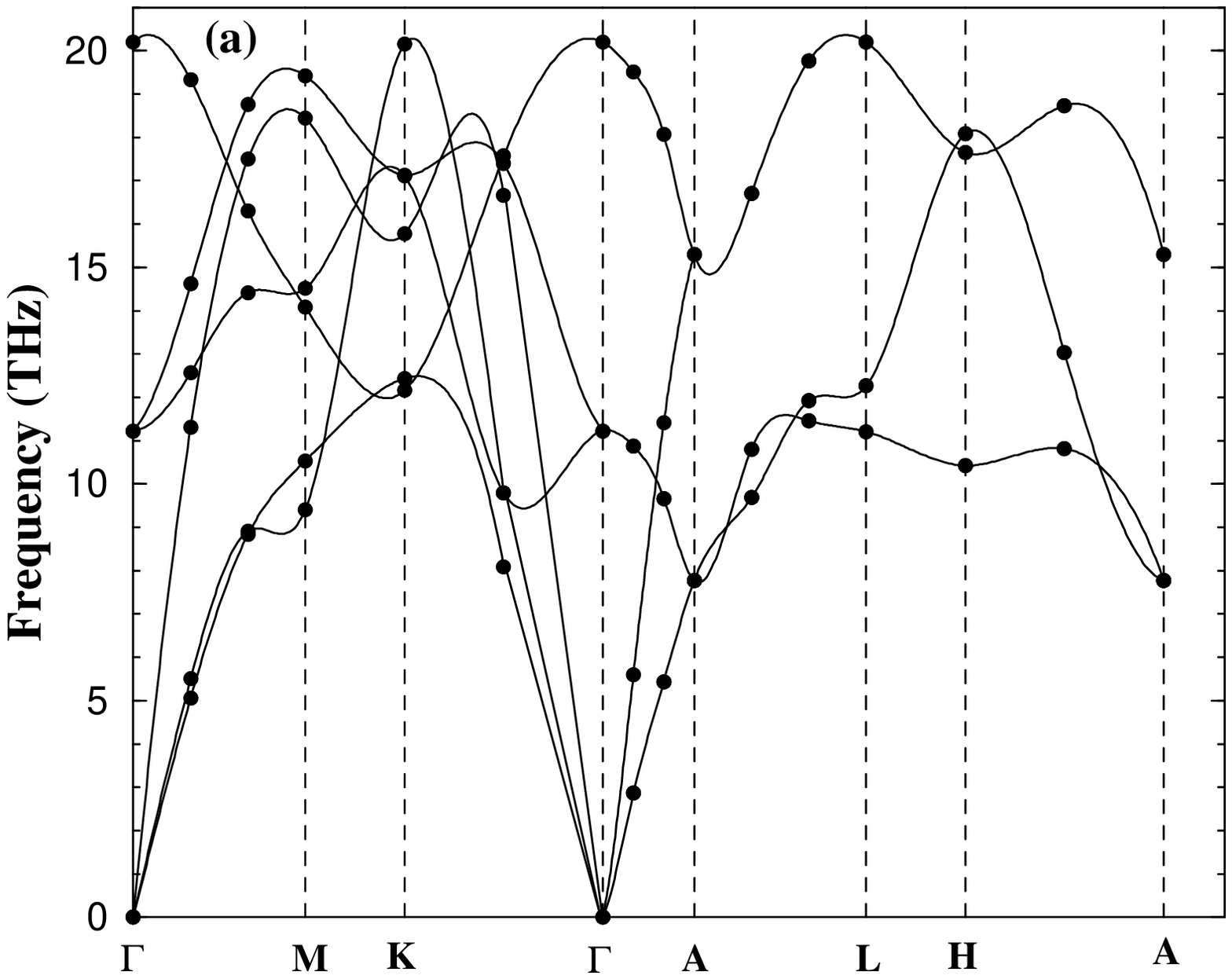}} 
\resizebox{!}{2.5in}{\includegraphics{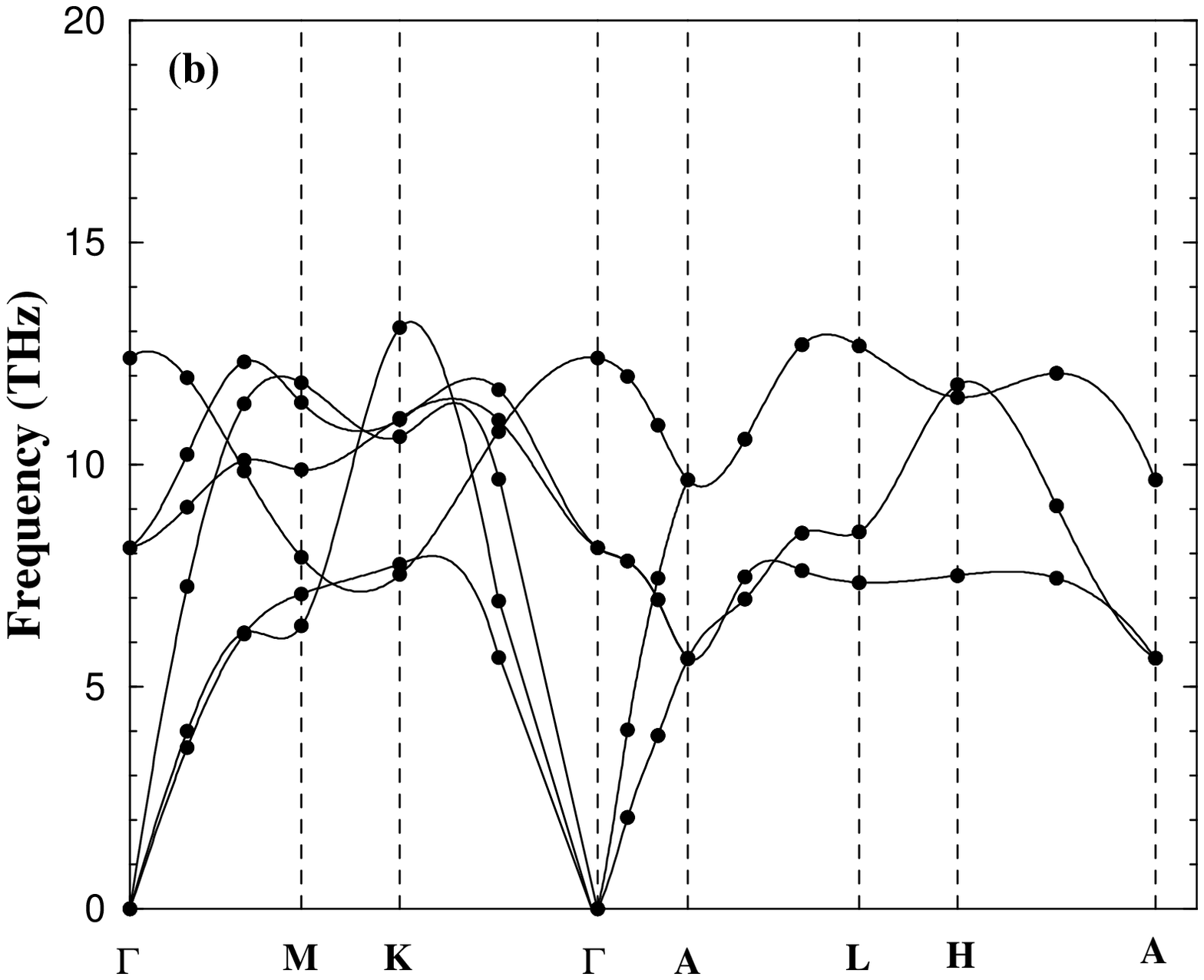}}
\caption[]{ Phonon frequencies of hcp Fe calculated via FP-LMTO-LR  method for  two different
lattice parameters: (a) 4.0 and (b) 4.4 a.u., and  the ideal c/a ratio, $\sqrt{8/3}$. 
The solid circles denote the calculated frequencies and the solid lines
represent spline fits through these calculated values.}
\label{fig2}
\end{figure}

We used the linear response code of Savrasov \cite{savrasov1,savrasov2} with a triple-$\kappa$ LMTO basis
set. The dynamical matrix 
was generated for 28 phonon wave vectors in the irreducible BZ, corresponding to a mesh of (6,6,6) reciprocal
lattice divisions. The BZ integration for the dynamical matrix was done for a mesh of
(12,12,12) reciprocal lattice divisions, and that for the electron-phonon (Hopfield) matrix
was done for a (24,24,24) mesh. 
 The calculated phonon spectra for two lattice parameters, 4.4 a.u. and 4.0 a.u., are
shown in Fig.\ref{fig2}. Our results are in reasonable agreement  with  a recent density functional calculation 
of Alf\'{e} {\it et al.} \cite{alfe}   These authors use the same GGA as  is used in our calculation
 (GGA1 \cite{PW1}) and the small
displacement method \cite{kresse} to obtain  the force constant matrix. In
 Fig.3 of their paper the phonon spectra for two volumes
8.67 \AA$^3$ and 6.97 \AA$^3$ are shown. The corresponding lattice parameters, 4.36 a.u. and 4.05 a.u.,
 are close to the values for which our calculated  phonon dispersions curves are shown in Fig. \ref{fig2}.
The phonon frequencies at the $\Gamma$ and  A points agree  remarkably  well.  Some differences
appear at symmetry points  K and  M. Such differences are also present between the results
of Alf\'{e} {\it et al.} \cite{alfe} and those obtained by S\"{o}derland {\it et al.}       using a 
generalized pseudopotential parameterization \cite{soderland} of FP-LMTO calculations. 
 The differences between our LR  results and those of 
Alf\'{e} {\it et al.} \cite{alfe} are not large enough to cause significant 
differences in thermal properties and
electron-phonon coupling.
 The smooth solid lines in Fig. \ref{fig2} correspond to spline fits to the calculated 
frequencies (solid circles).
Due to the small number of calculated frequencies the shapes of the lines presumably
representing the bands at the zone
boundaries could be incorrect.   
 The  connections of the calculated points with lines and 
band crossings in Fig. \ref{fig2} were determined by examining the phonon eigenvectors.
However,
the number of {\bf q}-points considered along each symmetry direction was at most four and  often three
or less.
No intermediate {\bf q}-points along the  K - M and  L - H were among the mesh of
{\bf q}-points for which the dynamical matrix was calculated. Thus 
 the possibility of   errors  in band crossing cannot be ruled out.

The dispersion curves at various pressures are similar, except for an
overall scale factor, essentially representing the gradual broadening of the bands with increasing pressure.
This is reflected in Fig. \ref{fig2} and also in the phonon density of states for various lattice parameters 
shown in part (b) of  Fig. \ref{fig4}. For the smallest lattice parameter considered by us  
the upper band edge lies around 670 cm$^{-1}$ or 20 THz (Fig. \ref{fig2}
and Fig. \ref{fig4} (b)).

We have computed both  the Eliashberg spectral function
\begin{equation}
\alpha^2F(\omega) =\frac{1}{N(0)}\sum_{{\bf k},{\bf k'},ij,\nu} |g_{{\bf k},{\bf k'}}^{ij,\nu}|^2
\delta(\varepsilon_{{\bf k}}^i) \delta(\varepsilon_{{\bf k'}}^j) \delta(\omega - 
\omega_{{\bf k}-{\bf k'}}^\nu)\;,
\end{equation}
and the transport Eliashberg function \cite{savrasov2,allen71}
\begin{eqnarray}
\alpha_{tr}^{2}F&  = & \frac{1}{2 N(0)\langle v_{F}^2\rangle}\sum_{{\bf k},{\bf k'},ij,\nu}
|g_{{\bf k},{\bf k'}}^{ij,\nu}|^2
   \left(\vec{v}_{F}({\bf k})-\vec{v}_{F}({\bf k'})\right)^2 \nonumber \\
 &  &  \times \delta(\varepsilon_{{\bf k}}^i) \delta(\varepsilon_{{\bf k'}}^j) \delta(\omega -
\omega_{{\bf k}-{\bf k'}}^\nu)\;,
\end{eqnarray}
where the subscript $F$ denotes  the Fermi surface, the angular brackets denote the  Fermi surface average,
$\vec{v}_{F}$ denotes the  Fermi surface velocity, $g_{{\bf k},{\bf k'}}^{ij,\nu}$ is 
the electron-phonon matrix element, with $\nu$
being the phonon polarization index and ${\bf k},{\bf k'}$ representing
electron wave vectors with band indices $i$, and $j$, respectively.

\begin{figure*}
\resizebox{!}{5.0in}{\includegraphics{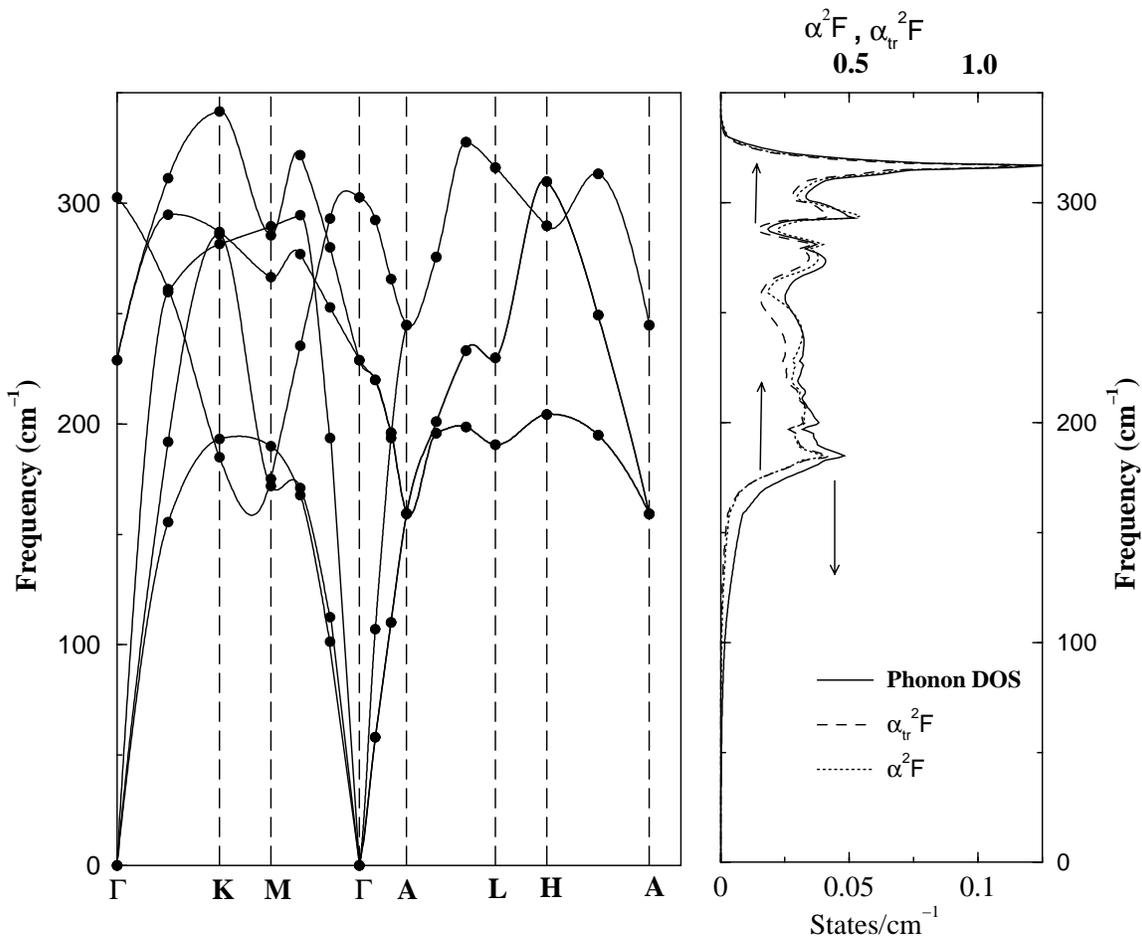}}
\caption[]{ Phonon spectrum, density of states and the Eliashberg spectral function
$\alpha^2F$ and the transport Eliashberg function $\alpha_{tr}^{2,xx}F$ for hcp Fe at the
lattice parameter 4.6 a.u. (c/a=$\sqrt{8/3}$). The equilibrium (minimum energy)
 lattice parameter is 4.615 a.u.}
\label{fig3}
\end{figure*}

\begin{figure}
\resizebox{!}{2.5in}{\includegraphics{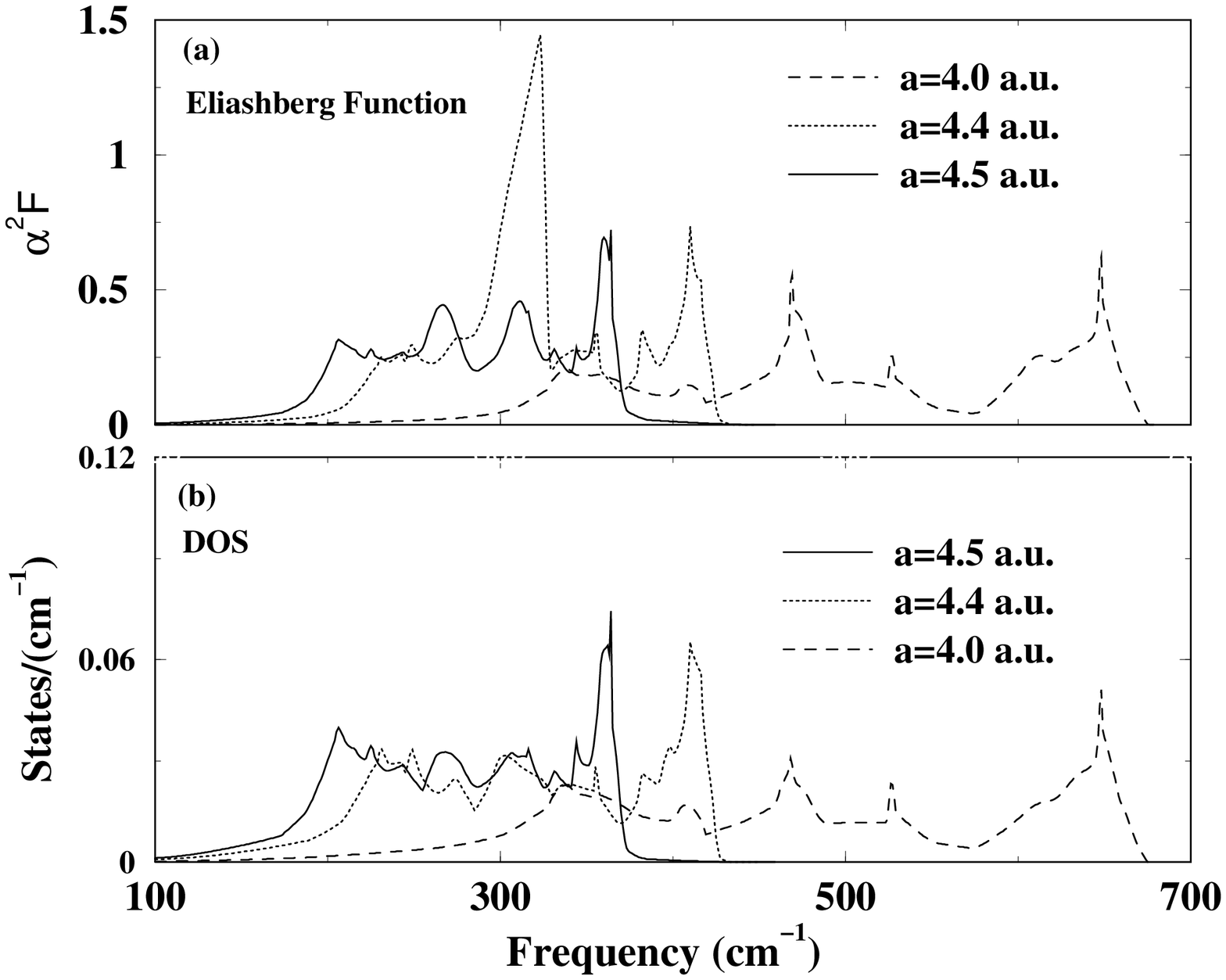}}
\caption[]{ Phonon density of states  and the Eliashberg function for hcp Fe for three different lattice parameters
(c/a=$\sqrt{8/3}$).}
\label{fig4}
\end{figure}

For most of the lattice parameters considered by us  the Eliashberg spectral  function 
$\alpha^2F$  and the transport Eliashberg function  $\alpha_{tr}^2F$  both 
follow the same frequency variation as the phonon density of states.
 In Fig.\ref{fig3} we show the phonon density of states and the two  Eliashberg functions
together with the phonon dispersions for the lattice parameter 4.6 a.u., close to the 
equilibrium value of 4.615 a.u.

 Some deviations in the frequency dependence of the $\alpha^2F$ function
from that of the phonon density of states appear at higher pressure. The deviation is 
most pronounced between 
the lattice parameters 4.4 and 4.2 a.u. in our calculation.  In Fig.\ref{fig4} we show the Eliashberg
spectral functions and the phonon density of states for three different lattice parameters.
The peaks in the  calculated phonon density of states  at ambient pressure (lattice parameter $\sim$ 4.6 a.u.
in our calculations)
are at 190 cm$^{-1}$ (24 meV) and 315 cm$^{-1}$ (39 meV), 
and these agree very well with the results from neutron scattering
experiments\cite{dos-neutron}  as well as with recently reported  results, obtained from the
measured energy spectra of inelastic nuclear absorption\cite{recent-dos}.  The peak positions in the
calculated results for higher pressures are at somewhat lower frequencies 
(by about 5 meV, which is within the experimental resolution) 
than those from the inelastic nuclear absorption experiment\cite{recent-dos}.
However, such  differences between the calculated and measured frequencies are common, given the
 difference between the experimental and theoretical values of the lattice parameters at
various  pressures. 

 The Hopfield parameter $\eta$ ($N(0)\langle I^2\rangle$),
 which is the electronic part of the electron-phonon coupling,
 shows an above average increase between the lattice parameters 4.5 and 4.4 a.u.,
due to an increased coupling for the longitudinal acoustic phonons with
wave vectors around the middle of  the $\Gamma$- K symmetry line.
This increased (above average) coupling is found to persist up to at least 4.2 a.u., but diminishes to normal
(average) value around the lattice parameter of 4.0 a.u., where the pressure-stiffening of the
lattice vibrations reduces the electron-phonon coupling and superconductivity disappears.
The general trend is as follows: the Hopfield parameter $\eta$ grows steadily with increasing pressure
with a rapid change between the lattice parameters 4.5 and 4.4 a.u. The phonon frequencies  move upward
with increasing pressure, with no phonon branches showing any softening. 
However, between lattice parameters 4.6 a.u. and 4.4 a.u. (perhaps 4.3 a.u.)
 the increase in the Hopfield parameter
dominates the change in the electron-phonon coupling parameter $\lambda_{ph} = \eta/M\langle\omega^2\rangle$.
  In this range $\lambda_{ph}$ increases 
despite a decrease in $N(0)$ and  an increase in $\langle\omega^2\rangle$. 
Below 4.3 a.u. lattice vibrations stiffen rapidly,
lowering the value of $\lambda_{ph}$. In Table \ref{table2} we summarize our results for
the pressure-dependence of the phonon properties and electron-phonon coupling. Since 
 the Hopfield parameter $\eta$ is often calculated using the rigid muffin-tin (RMT) approximation of 
Gaspari and Gyorffy \cite{RMT},
in Table \ref{table2} we have also presented the  results for $\eta$ obtained via  the LMTO-ASA
implementation of RMT (rigid atomic sphere or RSA) as given by Gl\"{o}tzel \etal \cite{glotzel} 
and Skriver and Mertig \cite{skriver}. These  values are in agreement with those given by Mazin 
\etal \cite{mazin1}, but differ significantly from the results of Jarlborg \cite{jarlborg} (judging from
 the quoted
values of $\lambda_{ph}$ and the Debye frequencies).
Our results indicate that depending on the lattice parameter the RMT/RSA 
approximation underestimates  the Hopfield parameter by 15-45\%. 
Also the variation of $\eta$ with lattice parameter in  the RMT approximation is much smoother than 
 in the linear response calculation, as the former fails to capture the above average increase
around the lattice parameter 4.4 a.u.
In Table \ref{table2} we have presented the results for lattice parameter 4.7 a.u. merely for comparison
with other lattice parameters, and not for comparison with experiment. The strong electron-phonon
coupling (stronger than that at $a=$4.6 a.u.) is of no  experimental consequence since,
(i) at this lattice parameter the system is at a negative pressure, not accessed by experiment; and
(ii) our theoretical calculations show that at this expanded volume the system is
most  likely antiferromagnetic.
\begin{table*}
\caption{
Hopfield parameters from the linear
response calculation $\eta$ and the rigid muffin-tin (atomic sphere) approximation $\eta$ (RMT/RAS), 
 mean square electron-ion matrix element $\langle I^2\rangle$, calculated average plasma frequencies $\omega_{pl}$, logarithmic
average phonon frequencies $\omega_{\ln}$,  cutoff frequencies $\omega_c$,
Coulomb pseudopentials for Eliashberg equation ($\mu^*(\omega_c)$)  
and McMillan formula ($\mu^*_{\text{ln}}$); 
electron-phonon coupling parameters $\lambda_{ep}$,
 calculated critical temperatures 
($T_c^{\text{calc}}$) and superconducting gaps ($\Delta_0$)
from the solution of the Eliashberg equations (\ref{eliash})
and the critical temperatures  from the McMillan formula (\ref{mcm})($T_c^{\text{McM}}$)
for various lattice parameters $a$. 
}
\label{table2}
\begin{ruledtabular}
\begin{tabular}{lrccccccc}
$a$ & $a_{\text B}$  & 4.0   & 4.2   & 4.4   & 4.5   & 4.6   & 4.7\\
$\eta$& Ry/bohr$^2$  & 0.268 & 0.368 & 0.229 & 0.139 & 0.111 & 0.099\\
$\eta$ (RMT/RAS)& Ry/bohr$^2$ 
                     & 0.214 & 0.167 & 0.124 & 0.108 & 0.095 & 0.088\\
$\langle I^2\rangle$ & (Ry/bohr)$^2$ & 0.056 & 0.063 & 0.032 & 0.018 & 0.013 & 0.010 \\
$\omega_{pl}$ & eV & 10.30 & 8.82 & 7.68 & 7.21 & 6.78 & 6.40\\
\hline
     &     K         & 640   & 542   & 439   & 372   & 336   &  295 \\
\rb{$\omega_{\text{ln}}$} & cm$^{-1}$
                     & 445   & 376   & 305   & 258   & 233   &  205 \\
\hline
$\omega_c$&cm$^{-1}$ & 7000  & 6000  & 4600  & 4600  & 4600  &  4490  \\
$\mu^*(\omega_c)$&   & 0.224 & 0.224 & 0.218 & 0.221 & 0.224 & 0.226 \\
$\mu^*(\omega_{\text{ln }})$&  
                     & 0.139 & 0.138 & 0.137 & 0.135 & 0.134 & 0.133 \\
$\lambda_{ph}$  &    & 0.277 & 0.570  & 0.538 & 0.434 & 0.431 & 0.508\\
$T_c^{\text{McM}}$& K& $<0.01$& 6.37 & 4.06 & 1.06 & 0.94 & 2.21 \\
$T_c^{\text{calc}}$& K& $5\cdot 10^{-7}$& 4.52 & 3.11 & 0.83 & 0.66 & 1.73 \\
$\Delta_0$& cm$^{-1}$& $< 10^{-6}$& 7.38 & 4.63 & 1.28 & 0.99 & 2.54 \\
$\Delta_0/k_BT_c^{\text{calc}}$ & & & 2.35 & 2.15 & 2.21 & 2.14 & 2.30\\
\end{tabular}
\end{ruledtabular}
\end{table*} 

\section{\protect\bigskip Critical temperature
}

\subsection{General relations}
 The linearized Eliashberg equations at the superconducting transition temperature $T_{c}$
 of an isotropic system  are (see, e.g., Ref. \onlinecite{allen-mitro}): 
\begin{eqnarray}
Z(i\omega _{n}) &=&1+\frac{\pi T_{c}}{\omega _{n}}\sum_{n^{\prime
}}W_{+}(n-n^{\prime })\mbox{sign}(n^{\prime }),  \label{eliash} \\
Z(i\omega _{n})\Delta (i\omega _{n}) &=&\pi T_{c}\sum_{n^{\prime }}^{\left|
\omega _{n}\right| \ll \omega _{c}}W_{-}(n-n^{\prime })\frac{\Delta (i\omega
_{n^{\prime }})}{\left| \omega _{n^{\prime }}\right| },  \notag
\end{eqnarray}%
where $\omega _{n}=\pi T_{c}(2n+1)$ is a Matsubara frequency, 
$\Delta (i\omega _{n})$ is an order parameter and $Z(i\omega _{n})$ is
a renormalization factor. Interactions
\begin{equation*}
W_{+}(n-n^{\prime })=\lambda _{ph}(n-n^{\prime })+\lambda _{sf}(n-n^{\prime
})+\delta _{nn^{\prime }}(\gamma _{n}+\gamma _{m}),
\end{equation*}
and
\begin{equation*}
W_{-}(n-n^{\prime })=\lambda _{ph}(n-n^{\prime })-\lambda _{sf}(n-n^{\prime
})-\mu ^{\ast }(\omega _{c})+\delta _{nn^{\prime }}(\gamma _{n}-\gamma _{m}),
\end{equation*}%
contain a phonon contribution
\begin{equation*}
\lambda _{ph}(n-n^{\prime })=\int_{0}^{\infty }\frac{d\omega ^{2}\alpha
^{2}(\omega )F(\omega )}{(\omega _{n}-\omega _{n^{\prime }})^{2}+\omega ^{2}}%
\;,
\end{equation*}%
where $\alpha ^{2}(\omega )F(\omega )$ is the so-called Eliashberg spectral
function, and  a contribution connected  with spin fluctuations%
\begin{equation*}
\lambda _{sf}(n-n^{\prime })=\int_{0}^{\infty }\frac{d\omega ^{2}P(\omega )}{%
(\omega _{n}-\omega _{n^{\prime }})^{2}+\omega ^{2}}\;.
\end{equation*}%
 $P(\omega )$  is  the spectral function of spin fluctuations, related
to the imaginary part of the  transversal spin susceptibility $\chi _{\pm }(\omega )$ as 
\begin{equation*}
P(\omega )=-\frac{1}{\pi }\left\langle \left| g_{\mathbf{kk}^{\prime
}}\right|^{2}\mathrm{Im}\chi _{\pm }(\mathbf{k,k}^{\prime },\omega
)\right\rangle\;,
\end{equation*}
 and $\mu ^{\ast }(\omega _{c})$ is the  screened Coulomb interaction

\begin{equation}
\mu ^{\ast }(\omega _{c})=\frac{\mu }{1+\mu \ln (E/\omega _{c})},  \label{mu}
\end{equation}%
where $\mu =\left\langle N(0)V_{c}\right\rangle _{FS}$ is  the Fermi surface averaged
Coulomb interaction.  E is a characteristic electron energy, $\omega
_{c}$  is a cut-off frequency, usually chosen ten times the
maximum phonon frequency: $\omega
_{c} \simeq 10\omega _{ph}^{\max }$.   $\gamma
_{n}=1/2\tau _{n},\gamma _{m}=1/2\tau _{m}$ are scattering rates for
nonmagnetic and magnetic impurities, respectively.

\subsection{Phonons only}

In order to compute $T_{c}$  we use the calculated Eliashberg spectral function
along with the
following procedure to determine the Coulomb pseudopotential $\mu ^{\ast }(\omega
_{c}).$   We start by assuming $\mu=0.5$.
  A value  greater than 0.5 for $\mu $ would lead to magnetic instability
(see, e.g. Ref. \onlinecite{scal}).
With
$E=\omega _{pl}$, the electron plasma frequency (see Ref. \onlinecite{pickett}),
\begin{equation*}
\mu ^{\ast }(\omega _{c})=\frac{0.5}{1+0.5\ln (\omega _{pl}/\omega _{c})}\;.
\end{equation*}
 Thus from the calculated phonon frequencies and plasma frequencies we obtain
$\mu ^{\ast }$ for all lattice parameters, with the cut-off frequency $\omega_{c}$ assumed to be
ten times the maximum phonon frequency. This procedure gives us the maximum possible values of
$\mu ^{\ast }(\omega _{c})$.

One of the most widely used expressions  for $T_{c}$ is given by the
Allen-Dynes form\cite{allen-mitro} of the McMillan formula (Eq.\ref{mcm}),
where
 the prefactor $\Theta_D/1.45$ is replaced by $\omega _{\ln  }^{ph}/1.2$. 
\begin{equation*}
\lambda _{ph}=2\int_{0}^{\infty }d\omega \alpha ^{2}(\omega )F(\omega
)/\omega
\end{equation*}
is  the electron-phonon  coupling constant, $\omega _{\ln  }^{ph}$ is
a logarithmically averaged characteristic phonon frequency
\begin{equation*}
\omega _{\ln  }^{ph}=\exp \left\{ \frac{2}{\lambda _{ph}}\int_{0}^{\infty }d\omega
\alpha ^{2}(\omega )F(\omega )\ln \omega /\omega\right\} ,
\end{equation*}%
and%
\begin{equation*}
\mu ^{\ast }\equiv \mu ^{\ast }(\omega _{\ln  }^{ph})=\frac{\mu ^{\ast
}(\omega _{c})}{1+\mu ^{\ast }(\omega _{c})\ln (\omega _{c}/\omega _{\ln 
}^{ph})}
\end{equation*}%
is the Coulomb pseudopotential at this frequency. Our calculations show that
for different plasma frequencies and characteristic phonon frequencies $\mu ^{\ast }$
for all lattice parameters lies in the range $0.13  -  0.14$, which is typical of
 conventional superconductors.

\begin{figure}
\resizebox{!}{2.5in}{\includegraphics{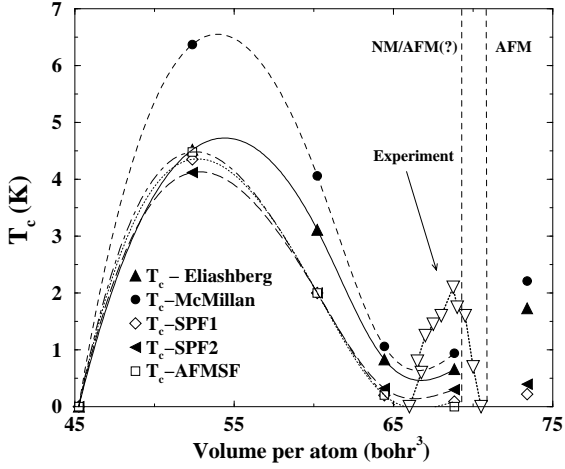}}
\caption[]{ Calculated transition temperatures for various volumes per atom. The experimental 
results are also shown. The experimental pressure versus $T_c$ results were transferred into volume versus $T_c$ results
using  the data from Mazin \etal\cite{mazin1} (see also Refs. \onlinecite{pvdata1,pvdata2}).
  The legends SPF1, SPF2, and AFMSF
are described in section IV. C. Dashed vertical lines show regions, where hcp Fe is known to be
antiferromagnetic (AFM), and where it is either nonmagnetic (NM) or antiferromagnetic (AFM).}
\label{fig5}
\end{figure}

 In Fig.\ref{fig5} we show the transition temperatures calculated as a function of volume per atom
using Eliashberg equations and the McMillan formula. The effects of ferromagnetic and
antiferromagnetic spin fluctuations, discussed in the following subsection, are also shown via
three additional curves. The symbols denote the calculated values of $T_c$ and the lines are spline fits
through the calculated values. 

\subsection{Contribution from spin fluctuations}

Superconducting transition temperatures calculated in the previous subsection are based on the
maximum possible estimates of the Coulomb pseudopotential $\mu ^{\ast }$. 
 Thus T$_c$, based on $s$-wave electron-phonon interaction only, cannot be less than 4.5 K, and
a value as high as 7-8 K is reasonable according to the linear response results. The highest
transition temperature obtained in the experiment\cite{nature1} is 2 K. A more important
difference between the calculated and the experimental results is the range of volume/pressure
over which superconductivity appears. The calculated range is much broader than the experimental
one (see Fig.\ref{fig5}). It is then natural to explore the effects of spin fluctuations on both,
the magnitude of $T_c$ and the pressure/volume range of the superconducting phase.
Since the calculation of the spin succeptibility is  rather
complicated, we restrict ourselves to  simple models of
\textit{ferromagnetic} and \textit{antiferromagnetic} spin fluctuations for an isotropic system,
as proposed by Mazin \etal \cite{mazin1}. 
 In a $T-$
matrix approximation for the  uniform electron  gas one can obtain 
the relation (see Refs. \onlinecite{BS,IVK}): 
\begin{equation*}
P(\omega )=N(0)\int_{0}^{2p_{F}}dq\frac{q}{2p_{F}^{2}}\left\{ -\frac{1}{\pi }
\mathrm{Im}\chi _{\pm }(q,\omega )\right\} ,
\end{equation*}%
where%
\begin{eqnarray*}
&  -\frac{1}{\pi }\mathrm{Im}\chi _{\pm }(q,\omega )\;\;\;\;  = \;\;\;\;   
\frac{I}{\pi}
\left[ \frac{\pi}{2} I N(0) \frac{\omega}{qv_{F}} \right] / &  \\
& \left[
  \left( 1 - I N(0) - I N(0) \frac{q^2}{12 p_F^2}
  \right)^2
 +  \left( \frac{\pi }{2}IN(0)\frac{\omega
}{qv_{F}}\right) \right]. &
\end{eqnarray*}
An integration of $P(\omega)$ (see section A.) leads to the spin fluctuation coupling parameter
\begin{equation}
\lambda _{sf}=\alpha N(0)I\ln \frac{1}{1-N(0)I},  \label{lsf}
\end{equation}%
where the constant $\alpha$ is of  order   unity. One can define
\begin{eqnarray}
\omega^{sf} _{\ln } & = & \exp \frac{2}{\lambda _{sf}}\int_{0}^{\infty }d\omega P(\omega
)\ln \omega /\omega \nonumber \\
 & & \approx (0.8)\frac{[1-IN(0)]}{I N(0)}p_{F}v_{F}  \label{sflog}
\end{eqnarray}
as a characteristic spin fluctuation frequency,
which should vanish near  the magnetic phase transition. $p_{F}$, and $v_{F}$ are the Fermi momentum and
velocity, respectively. The product $p_{F}v_{F}$ can be replaced by 
 $2E_F$ and estimated from the location of the
Fermi energy with respect to the bottom of the band.

If we  resort to the approximation
\begin{equation*}
P(\omega )=(\lambda _{sf}\omega _{sf}/2)\delta (\omega -\omega _{sf})\;,
\end{equation*}%
then for  $\omega _{sf}\gg \omega _{ph}$ we obtain  an extension of the McMillan
formula, similar to the one used in Ref. \onlinecite{mazin1}
\begin{equation}
T_{c}=\frac{\omega _{\ln  }^{ph}}{1.2}\exp \left\{ -\frac{1.04(1+\lambda
_{ph}+\lambda _{sf})}{\lambda _{ph}-\lambda _{sf}-\mu ^{\ast
}[1+0.62(\lambda _{ph}+\lambda _{sf})]}\right\} .  \label{maz}
\end{equation}%
In reality the spectrum $P(\omega )$ is distributed from zero up to
electronic energies. Near the phase transition the characteristic frequency
is comparable to  the characteristic phonon frequencies. An appropriate
treatment of the broadness of the spectrum $P(\omega )$ 
leads to the replacement of the $\omega _{\ln  }^{ph}$ in  the above expression by

\begin{equation}
\omega =\omega^{sf} _{\ln }(\omega _{\ln  }^{ph}/\omega^{sf} _{\ln })^{\nu },
\label{omega}
\end{equation}%
with the  exponent $\nu$ (see Ref. \onlinecite{IVK}) given by
\begin{equation*}
\nu =\frac{\lambda _{ph}^{2}}{(\lambda _{ph}-\lambda _{sf})\left[ \lambda
_{ph}-\lambda _{sf}+\frac{\lambda _{ph}\lambda _{sf}}{1+\lambda
_{ph}+\lambda _{sf}}\ln [\omega _{\ln  }^{ph}/\omega _{sf}]\right] }.
\end{equation*}

 In the uniform electron gas approximation  the constant $\alpha$ in Eq. (\ref{lsf}) is of the order of unity.  But
  such a high
value of $\alpha$ in our calculation
 would cause the critical temperatures to  vanish for all lattice parameters ($\lambda _{ph}<\lambda _{sf}$).
 It is  evident that the uniform electron gas  approximation would be inappropriate 
 for a transition metal like iron.  Hence we use the following approach: we  consider  $\alpha $ as a fitting parameter to
get a  $T_{c}=$2 K, the experimental value, for the lattice parameter 4.4 a.u. (volume per atom
$\sim 60$ a.u. The two sets of $T_c$
 versus lattice parameter results obtained this way are shown in Fig.(\ref{fig5}) 
and are labeled as SPF1 and SPF2, respectively (the lowermost curves). In particular, 
SPF1 refers to the case where Eq. (\ref{maz}) is used with $\lambda_{sf}$ given by
Eq. (\ref{lsf}); and SPF2 refers to the case  where $\omega$ from Eq. (\ref{omega}) replaces $\omega^{ph}_{\ln }$ in Eq.(\ref{maz}),
with $\lambda_{sf}$ still given by Eq. (\ref{lsf}). In Table \ref{table3}, the spin fluctuation coupling
parameters associated with the results SPF1 and SPF2 in Fig. \ref{fig5} are  labeled as
$\lambda_{sf1}$ and $\lambda_{sf2}$, respectively. 
 The values of the parameter $\alpha$ (see Eq.(\ref{lsf})) for the two cases SPF1 and SPF2 are 0.101  and
0.029, respectively. 
Calculated transition temperatures for the two models SPF1 and SPF2
are denoted by $T^{sf}_{c1}$ and  $T^{sf}_{c2}$ in Table \ref{table3}, 
which also shows the characteristic spin fluctuation frequencies $\omega^{sf}_{\ln }$
for various lattice parameters.

\begin{table*}
\caption{Spin fluctuation effects: ferromagnetic spin fluctuation coupling parameters
$\lambda_{sf1}, \lambda_{sf2}$ (see text in  section IV. C.  for details); antiferromagnetic spin fluctuation
coupling parameter $\lambda^{af}_{sf}$, the characteristic spin fluctuation frequency
$\omega^{sf}_{ln }$, and the corresponding critical temperatures $T^{sf}_{c1}, T^{sf}_{c2}$,
and $T^{af}_{c,sf}$ (see text in  section IV. C. for details).}
\label{table3}
\begin{ruledtabular}
\begin{tabular}{llcccccc}
$a$ & $a_{\text B}$  & 4.0   & 4.2   & 4.4   & 4.5   & 4.6   & 4.7\\
\hline\\
$\omega^{sf}_{ln }$     &     eV        & 33.48 & 22.56 & 14.80 & 11.39 & 8.59 & 5.67\\
$\lambda _{sf1}$ & & 0.0155 & 0.024 & 0.038 & 0.048 & 0.062 & 0.0886 \\
$\lambda _{sf2}$ &  &0.0044 & 0.0069 & 0.011 & 0.0139 & 0.018 & 0.025 \\
$\lambda _{sf}^{af}$ &  & 0.0036 & 0.0057 & 0.011 & 0.019 & 0.05 &  \\
$T_{c1}^{sf}$ & K  & 0.0005 & 4.35  & 2 & 0.202 & 0.091 & 0.224 \\
$T_{c2}^{sf}$ & K  & 0.0019 & 4.12  & 2 & 0.324 & 0.303 & 0.398 \\
$T_{c}^{af}$ & K  & 0.0023 & 4.48 & 2 & 0.206 & 0.004 & 0%
\end{tabular}
\end{ruledtabular}
\end{table*}

For \textit{antiferromagnetic} spin fluctuations the spin susceptibility has
a maximum at $\mathbf{q\rightarrow Q}^{\ast }$, and the averaging over the
Fermi surface leads to
\begin{equation*}
\lambda _{sf}\approx \frac{\chi (\mathbf{q\rightarrow Q}^{\ast })I}{1-\chi (%
\mathbf{q\rightarrow Q}^{\ast })I}.
\end{equation*}%
If, according to  Mazin \etal \cite{mazin1}, we suppose $\chi (\mathbf{q\rightarrow Q}
^{\ast })=bN(0)$, then

\begin{equation}
\lambda _{sf}^{af}=\frac{\alpha ^{\prime }bN(0)I}{1-bN(0)I},
\end{equation}%
Parameter $b$ can be estimated from the condition of the antiferromagnetic
instability $bN(0)I\rightarrow 1.$  This leads to $b\lesssim 1.5$ which is
close to the value in Ref. \onlinecite{mazin1}. Taking this value and using $%
\alpha ^{\prime }$ as a fitting parameter we obtain the result plotted in Fig.
\ref{fig5}.  $\alpha^\prime = 0.0032$ reduces the maximum $T_c$ (at 4.4 a.u.) to the
experimental value, 2 K (for
simplicity we used Eq.(\ref{maz}), with $\omega_{ln }^{ph}$ replaced by $\omega$
given by Eqns. (\ref{sflog}) and (\ref{omega}), and $\lambda_{sf}^{af}$ replacing $\lambda_{sf}$). 
The corresponding results are plotted in Fig.(\ref{fig5}) and are labeled as AFMSF. 
The values of $T_c$ and spin fluctuation coupling parameters are also shown in
Table \ref{table3}, labeled as $T_c^{af}$ and $\lambda_{sf}^{af}$, respectively.
It is evident that the volume dependence of  $T_{c}$ obtained this way
is  very similar  for ferromagnetic  and antiferromagnetic spin fluctuations.

\subsection{Magnetic impurities}
\begin{figure}
\resizebox{!}{2.5in}{\includegraphics{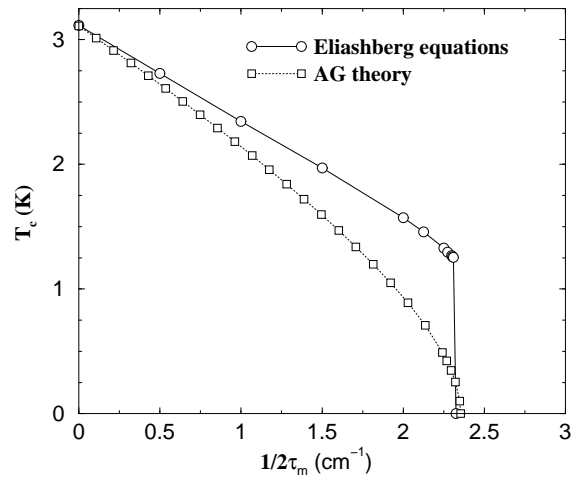}}
\caption[]{ Variation of the transition temperature with the scattering rate of magnetic impurities
 1/$2\tau_m$ for lattice parameter a =4.4 a.u. 
 The legend AG stands for solution of
the  Abrikosov-Gor'kov expression given by Eq.(\ref{AG}). }
\label{fig6}
\end{figure}

 A lowering of the critical temperature could also be  caused by the
presence of magnetic impurities. It is well-known  that the nonmagnetic
impurities  cancel out from the Eliashberg equations (Anderson theorem \cite{anderson} ), while the magnetic
ones lead the pair-breaking effects\cite{pair-break}. The central idea  is that near a magnetic
transition spin- ordered clusters appear,  and these  can scatter
electrons very effectively. We have calculated  the effect of such impurities on
the critical temperature for the lattice parameter $a=4.4$ a.u. (see, Fig.
\ref{fig6})  by considering various different scattering rates $1/2\tau _{m}$ in 
the Eliashberg Eq.(\ref{eliash}).
For  comparison we have also calculated the change in  $T_{c}$ by using
the renormalized Abrikosov-Gor'kov (AG) expression (see e.g.,  section 15 in Ref.
 \onlinecite{allen-mitro})
\begin{equation}
\ln (T_{c0}/T_{c})\simeq \psi \left[ 1/2+(1/2\tau _{m})\pi T_{c}(1+\lambda
_{ep})\right] -\psi (1/2),
\label{AG}
\end{equation}
where $T_{c0}$ is the critical temperature without magnetic impurities.
$\psi (x) $ is the digamma function and $\psi (1/2)$ is related to the Euler constant $\gamma$ as
$\psi (1/2) = -\gamma - 2\ln 2$.
The difference  between the Eliashberg and the AG results
is due to  the rather broad phonon spectrum which necessitates
 appropriate treatment of  strong coupling effects.

In order to reduce the critical temperature  at the lattice parameter $a=4.4$ a.u. to
the experimental value of  2 K it is  sufficient to
  assume a scattering rate $1/\tau _{m}\approx 3.0$ cm$^{-1}$.  With the
calculated average Fermi velocity $v_{F}\simeq 2.547\times 10^7$ cm/sec, this
yields a mean free path $l\simeq 28.3\times 10^{-5} $ cm. The closer the magnetic
instability, the larger  is the probability (rate) of  magnetic scattering, 
 which leads  to more enhanced suppression of $T_{c}$.

\subsection{p-wave pairing}

Magnetic  ordering (as well as an external magnetic field) favors
the triplet p-wave pairing, similar to  that found in superfluid $^{3}$He.
In order to  estimate  $T_c$ for p-wave pairing
 we adopt the
following simplified approach.  We consider the extension of
 Eq. (\ref{eliash}) for  the $l^{th}$ spherical harmonic
channel  \cite{allen-mitro}:

\begin{eqnarray}
Z(i\omega _{n}) &=&1+\frac{\pi T_{c}}{\omega _{n}}\sum_{n^{\prime
}}W_{+}^{(0)}(n-n^{\prime })\mbox{sign}(n^{\prime }),  \label{p} \\
\mathbf{d}^{(l)}(i\omega _{n}) &=&\pi T_{c}\sum_{n^{\prime }}^{\left| \omega
_{n}\right| \ll \omega _{c}}W_{-}^{(l)}(n-n^{\prime })\frac{\mathbf{d}%
^{\left( l\right) }(i\omega _{n^{\prime }})}{Z(i\omega _{n^{\prime }})\left|
\omega _{n^{\prime }}\right| },  \notag
\end{eqnarray}%
where $\mathbf{d}^{(l)}$ for $l=1$ is the  p-wave order parameter, and

\begin{equation*}
W_{+}^{(l)}(n-n^{\prime })=\lambda^l _{ph}(n-n^{\prime })+\lambda^l
_{sf}(n-n^{\prime })+\delta_{l0}\delta _{nn^{\prime }}(\gamma _{n}+\gamma _{m}),
\end{equation*}%
\begin{equation*}
W_{-}^{(l)}(n-n^{\prime })=\lambda_{ph}^{l}(n-n^{\prime })+(-1)^l \lambda
_{sf}^{l}(n-n^{\prime })+ \delta_{l0} X \; ,
\end{equation*}%
where $X = -\mu ^{\ast }(\omega _{c})+\delta _{nn^{\prime }}(\gamma _{n}-\gamma _{m})$.
The kernel $W$ with a general index $l$ is defined as 
 the Fermi surface (FS) average

\begin{equation*}
W^{(l)}=\left\langle \mathbf{d}^{(l)}W(\mathbf{k,k}^{\prime },n-n)\mathbf{d}%
^{(l)}\right\rangle _{\mathbf{k,k}^{\prime }\in FS}/\left\langle \left|
\mathbf{d}^{(l)}\right| ^{2}\right\rangle
\end{equation*}%
 of the $l^{th}$ harmonic of the momentum-dependent interaction $W(\mathbf{k,k}%
^{\prime },n-n)$, while
\begin{equation*}
W^{(0)}=\left\langle W(\mathbf{k,k}^{\prime },n-n)\right\rangle _{\mathbf{k,k%
}^{\prime }\in FS}.
\end{equation*}%
We  assume that the Coulomb interaction and impurity scattering are isotropic.
The simplest approximation  then is to use $W^{(1)}=gW^{(0)}$, where the parameter $g$
describes the anisotropy of the interaction (see, e.g., Refs. \onlinecite{dg,jcd}).
A difference of the factor $g$ from unity leads to strong pair-breaking
effects. In general (see., e.g. Ref. \onlinecite{allen-mitro}),  for $l=1$ $\mathbf{d}^{(l)}\approx
\mathbf{v}_{F}$, the first odd Fermi surface harmonic\cite{allen76}. 
In this case $g=\lambda _{ph}^1/\lambda _{ph}^0 = \lambda _{ph}^{(in)}/\lambda _{ph}$ (see
notations in 
Refs.\onlinecite{allen71},  \onlinecite{savrasov2}). The
phonon constant $\lambda _{ph}^{(in)}$ is relevant to the transport Boltzmann equation (see, e.g.,
  Ref. \onlinecite{savrasov2}).  From the linear response calculation we obtain $g= \lambda _{ph}^{(in)}/\lambda _{ph}$
 for all lattice parameters. For a=4.4 a.u., we obtain g=0.238.

With the  assumption $W^{(1)}=gW^{(0)}$ it is possible to estimate $T_c$ from a McMillan-like 
formula.
An expression for the critical temperature can be written in a form similar
to  that used by Mazin {\it et al.} \cite{mazin1}:

\begin{eqnarray}
T_{c} & = & \frac{\omega}{1.2}\exp \left\{ -\frac{1+\lambda^0_{ph}+\lambda^0_{sf}}{
\lambda^1_{ph}+\lambda^1_{sf}}\right\} \nonumber\\
 & =  & \frac{\omega}{1.2}\exp \left\{ -\frac{1+\lambda _{ph}+\lambda _{sf}}{
g(\lambda_{ph}+\lambda _{sf})}\right\} ,  \label{p-mcm}
\end{eqnarray}
where $\omega $ is given by Eq. (\ref{omega}), and we have used the
relation: $
 \lambda^1_{sf} = g \lambda^0_{sf} = g \lambda_{sf}$.
Note that this equation is the same as equation (3.6) of Fay and Appel
\cite{fay_appel}, except that the term $\lambda^1_{ph}$ is absent from the
exponent in their expression for $T_c$ \cite{footnote}.  
A small value of the parameter $g$ 
 and a rather strong phonon contribution to the numerator  in the 
exponent in  Eq.(\ref{p-mcm}) lead to small values of $T_{c}\lesssim
10^{-2}K$ for the $p-$wave pairing in  contrast to the conclusion reached in Ref. 
\onlinecite{jarlborg}. The value $T_{c}\lesssim
10^{-2}K$ is similar to that obtained by Allen and Mitrovi\'{c} \cite{allen-mitro}
for $p$-wave superconductivity in Pd. If the assumption $W^{(1)}=gW^{(0)}$ is valid, then the
inclusion of antiferromagnetic spin fluctuations (replacing $\lambda_{sf}$ by $\lambda^{af}_{sf}$)
 would lead to similar results.

\section {Summary and Conclusions}

The state of hcp Fe under moderately high pressure ($\leq 60$ GPa) is  currently riddled with
controversial results that need to be  resolved and understood. The initial experimental result by
Shimizu \etal \cite{nature1} indicating superconductivity between 15 and 30 GPa has been verified recently
\cite{jaccard} for a pressure of 22.5 GPa, where the maximum $T_c$ of 2 K was observed. Experimental
measurements of Raman spectra are also reported \cite{merkel,goncharov}.  
The observation of a second Raman peak besides that due the Raman active $E_{2g}$ mode
in hcp Fe has been assigned to disorder induced phonon scattering \cite{merkel,goncharov}, 
 although, based on density functional calculation, the possibility of antiferromagnetic
order up to a pressure of approximately 60 GPa has also been suggested 
\cite{cohen2}.
 Normal state resistivity measurements by Jaccard \etal show a $T^{5/3}$ temperature dependence at low
temperature, which is predicted by a nearly ferromagnetic Fermi-liquid model. The earlier 
M\"{o}ssbauer study of hcp Fe under pressure had failed to reveal any local magnetic moment
\cite{cort,taylor}, and results of theoretical calculations are dependent on the treatment
of the exchange and correlation potentials used in the calculations. In view of this, we have adopted
the same approach as  Mazin \etal \cite{mazin1}: we assume nonmagnetic state under pressure and examine the 
electron-phonon coupling as a function of the  lattice parameter. 

The results of our study can be summarized as follows:\\
(i) The Hopfield parameter $\eta$ increases steadily with pressure, showing a wider
variation for the linear response calculation  than obtained via the RMT approximation (both
in our LMTO-RMT calculation and in that of Mazin \etal \cite{mazin1} ).
(ii) Below volumes $\sim $50 a.u. per atom (above estimated pressures $\sim$ 160 GPa) phonons
stiffen rapidly, bringing  the $T_c$ down (somewhat faster than what is suggested in Ref. \onlinecite{mazin1}).
(iii) $T_c$'s based on the $s$-wave  electron-phonon coupling, and maximum possible
estimates of $\mu ^{\ast }$  are  higher than the experimental values 
(iv) The range of volume where superconductivity appears is much broader in the calculations than
what is observed, in agreement with the  result of  Ref. \onlinecite{mazin1}.
(v) Inclusion of  ferromagnetic/antiferromagnetic spin fluctuations, and scattering from
magnetic impurities  can all  bring the calculated values of $T_c$ down to the range of observed
values, but cannot substantially improve the agreement between the calculated and the
experimental   pressure/volume range of the superconducting phase (Fig.\ref{fig5}).
(vi) A simplified treatment of $p$-wave pairing due to electron-phonon and spin fluctuation
interactions yields  a very small $T_c$ ($\leq$ 0.01 K), in contrast with  
 the claim made in Ref. \onlinecite{jarlborg}.

The role of impurities remains somewhat puzzling as well as of vital interest at present.
The initial results of Shimizu \etal \cite{nature1} showed very high normal state residual
resistivity ($\sim 40 \mu\Omega$cm), indicating the presence of substantial defects in the
samples studied. More recent measurements \cite{jaccard} with 
purer samples (with residual resistivity 50 times smaller) show the same maximum $T_c$ at the
same pressure. If the impurities in the earlier studied samples were magnetic this would
rule out electron-phonon $s$-wave coupling as the primary mechanism of superconductivity.
However, Jaccard \etal \cite{jaccard} also notice that superconductivity in their
samples is unusually sensitive to disorder, developing only  when the electronic mean free path
exceeds a threshold value. They find the normal state resistivity to be characteristic of a
nearly ferromagnetic metal, but our calculation of $p$-wave superconducting $T_c$ in the 
presence of ferromagnetic spin fluctuations and electron-phonon interactions yields  values less than
$10^{-2}$ K,  similar to that for Pd  obtained by Allen and Mitrovi\'{c} \cite{allen-mitro}.

Our calculations seem to rule out both $s$- and $p$-wave superconductivity in hcp Fe.
Electron-phonon mediated $s$-wave superconductivity should persist, in severe
disagreement with experiment, beyond pressures of 200 GPa
even in the presence of spin fluctuations.
The possibility of $d$-wave superconductivity mediated by antiferromagnetic spin fluctuations
remains to be explored. A $d$-wave superconductivity would be consistent with the observation \cite{jaccard}
that superconductivity in hcp Fe seems to be extreemly sensitive to disorder. Low temperature
specific heat measurements can further clarify this issue.

\begin{center}ACKNOWLEDGMENTS
\end{center}
SKB would like to thank S.Y. Savrasov and D.Y. Savrasov for helpful hints and discussions
related to the linear response code.

\end{document}